\documentclass{article}

\usepackage{arxiv}

\usepackage[utf8]{inputenc} 
\usepackage[T1]{fontenc}    
\usepackage{hyperref}       
\usepackage{url}            
\usepackage{booktabs}       
\usepackage{amsfonts}       
\usepackage{nicefrac}       
\usepackage{microtype}      
\usepackage{lipsum}
\usepackage{graphicx}
\usepackage{nameref}
\usepackage{makecell}
\usepackage{appendix}

\title{Stopwords in Technical Language Processing}

\author{
  Serhad Sarica \\
  Data-Driven Innovation Lab\\
  Singapore University of Technology and Design\\
  Singapore, 487372 \\
  \texttt{serhad\_sarica@mymail.sutd.edu.sg} \\
   \And
 Jianxi Luo \\
  Data-Driven Innovation Lab\\
  Singapore University of Technology and Design\\
  Singapore, 487372 \\
  \texttt{luo@sutd.edu.sg} \\
}

\begin{document}
\maketitle

\begin{abstract}
There are increasingly applications of natural language processing techniques for information retrieval, indexing and topic modelling in the engineering contexts. A standard component of such tasks is the removal of stopwords, which are uninformative components of the data. While researchers use readily available stopwords lists which are derived for general English language, the technical jargon of engineering fields contains their own highly frequent and uninformative words and there exists no standard stopwords list for technical language processing applications. Here we address this gap by rigorously identifying generic, insignificant, uninformative stopwords in engineering texts beyond the stopwords in general texts, based on the synthesis of alternative data-driven approaches, and curating a stopwords list ready for technical language processing applications.
\end{abstract}

\keywords{Stopwords \and Technical language \and Data-driven}

\section{Introduction}
\label{sec:intro}
Natural language processing (NLP) and text analysis have been growingly popular in engineering analytics \cite{Chang2015, Zhang2014, Grawe2017,Liu2020}. To ensure the accuracy and efficiency of such NLP tasks as indexing, topic modelling and information retrieval \cite{Blanchard2007, Gerlach2019, Tsz-WaiLo2005, Fox1989, Wilbur1992}, the uninformative words, often referred to as “stopwords”, need to be removed in the pre-processing step, in order to increase signal-to-noise ratio in the unstructured text data. Example stopwords include ”each”, ”about”, ”such” and ”the”. Stopwords often appear frequently in many different natural language documents or parts of the text in a document but carry little information about the part of the text they belong to.

The use of a standard stopword list, such as the one distributed with popular Natural Language Tool Kit (NLTK) \cite{bird2009natural} python package, for removal in data pre-processing has become an NLP standard in both research and industry. There have been efforts to identify stopwords from generic knowledge sources such as Brown Corpus \cite{Fox1989, Kucera1969}, 20 newsgroup corpus [6], books corpus \cite{Montemurro2010}, etc, and curate a generic stopword list for removal in NLP applications across fields. However, the technical language used in engineering or technical texts is different from layman languages and may use stopwords that are less prevalent in layperson languages. When it comes to engineering or technical text analysis, researchers and engineers either just adopt the readily available generic stopword lists for removal  \cite{Chang2015, Zhang2014, Grawe2017,Liu2020} leaving many noises in the data or identify additional stopwords in a manual, ad hoc or heuristic manner \cite{Blanchard2007, Sarica2020, Seki2005, Crow2004}. There exist no standard stopword list for technical language processing applications.

Here, we address this gap by rigorously identifying generic, insignificant, uninformative stopwords in engineering texts beyond the stopwords in general texts, based on the synthesis of alternative data-driven approaches. The resultant stopword list is statistically identified and human-evaluated. Researchers, analysts and engineers working on technology-related textual data and technical language analysis can directly apply it for denoising and filtering of their technical textual data, without conducting the manual and ad hoc discovery and removal of uninformative words by themselves.

\section{Our approach}
\label{sec:approach}
To identify stopwords in technical language texts, we statistically analyse the natural texts in patent documents which are descriptions of technologies at all levels. The patent database is vast and provides the most comprehensive coverage of technological domains. Specifically, our patent text corpus contains 781,156,082 tokens (words, bi-, tri- and four-grams) from 30,265,976 sentences of the titles and abstracts of 6,559,305 of utility patents in the complete USPTO patent database from 1976 to 31st December 2019 (access date: 23 March 2020). Non-technical design patents are excluded. Technical description fields are avoided because they include information on contexts, backgrounds and prior arts that may be non-relevant to the specific invention and repetitive, lead to statistical bias and increase computational requirements. We also avoided legal claim sections which are written in repetitive, disguising and legal terms.

In general text analysis for topic modelling or information retrieval, various statistical metrics, such as term frequency (TF) \cite{Tsz-WaiLo2005,Wilbur1992}, inverse-document frequency (IDF) \cite{Tsz-WaiLo2005}, term-frequency-inverse-document-frequency (TFIDF) \cite{Blanchard2007}, entropy \cite{Gerlach2019, Montemurro2010}, information content \cite{Gerlach2019}, information gain \cite{Makrehchi2008} and Kullback-Leibler (KL) divergence \cite{Tsz-WaiLo2005}, are employed to sort the words in a corpus \cite{Gerlach2019, Makrehchi2008}. Herein we use TF, TFIDF and information entropy to automatically identify candidate stopwords.

Furthermore, some of the technically significant terms such as “composite wall”, “driving motion” and “hose adapter” are statistically indistinguishable from such stopwords “be”, “and” and “for”, regardless of the statistic metrics for sorting. That is, automatic and data-driven methods by themselves are not accurate and reliable enough to return stopwords. Therefore, we also use a human-reliant step to further evaluate the automatically identified candidate stopwords and confirm a final set of stopwords which do not carry information on engineering and technology.

\begin{figure}[h]
  \centering
  \includegraphics[scale=0.50]{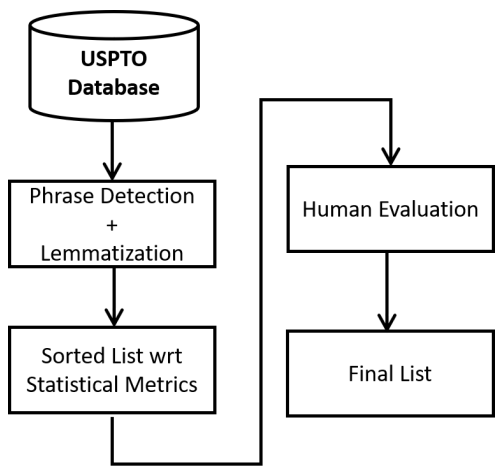}
  \caption{Overall procedure}
  \label{fig:fig1}
\end{figure}

In brief, the overall procedure as depicted in Figure \ref{fig:fig1} consists of three major steps: 1) basic pre-processing of the patent natural texts, including punctuation removal, lower-casing, phrase detection and lemmatization; 2) using multiple statistic metrics from NLP and information theory to identify a ranked list of candidate stopwords; 3) term-by-term evaluation by human experts on their insignificance for technical texts to confirm stopwords that are uninformative about engineering and technology. In the following, we describe implementation details of these three steps.

\section{Implementation}
\label{sec:implement}
\subsection{Pre-processing}
\label{sec:preprocess}
The patent texts in the corpus are first transformed into a line-sentence format, utilizing the sentence tokenization method in the NLTK, and normalized to lowercase letters to avoid additional vocabulary caused by lowercase/uppercase differences of the same words. The punctuation marks in sentences are removed except “-” and “/”. These two special characters are frequently used in word-tuples, such as “AC/DC” and “inter-link”, which can be regarded as a single term. The original raw texts are transformed into a collection of 30,265,976 sentences, including 796,953,246 unigrams.

Phrases are detected with the algorithm of Mikolov et al \cite{Mikolov2013} that finds words that frequently appear together, and in other contexts infrequently, by using a simple statistical method based on the count of words to give a score to each bigram such that:

\begin{equation}
score(w_{i},w_{j}) = \frac{(count(w_{i}w_{j})-\delta)|N|}{count(w_{i})count(w_{j})}
\end{equation}

where $count(w_{i}w_{j})$ is the count of $w_{i}$ and $w_{j}$ appearing together as bigrams in the collection of sentences and $count(w_{i})$ is the count of $w_{i}$ in the collection of sentences. $\delta$ is the discounting coefficient to prevent too many phrases consisting of very infrequent words, and set $\delta = 1$ to prevent having scores higher than 0 for phrases occurring less than twice. The term $N = \sum\limits_{t,p\in P}n(t,p)$   represents the total number of tokens in the patent database where $n(t,p)$ is the count of the term $t$ in the patent $p$. Bigrams with a score over a defined threshold $(T_{phrase})$ are considered as phrases and joined with a “\_” character in the corpus, to be treated as a single term. We run the phrasing algorithm of Mikolov et al. \cite{Mikolov2013} on the pre-processed corpus twice to detect n-grams, where n = [2,4]. The first run detects only bigrams by employing a higher threshold value $T_{phrase}^{1}$, while the second run can detect n-grams up to n = 4 by using a lower threshold value $T_{phrase}^{2}$ to enable combinations of bigrams. Via this procedure of repeating the phrasing process with decreasing threshold values of $T_{phrase}$, we detected phrases that appear more frequently in the first step using the higher threshold value, e.g., “autonomous vehicle”, and detected phrases that are comparatively less frequent in the second step using the lower threshold value, e.g., “autonomous vehicle platooning”. In this study, we used the best performing thresholds (5, 2.5) found in a previous study \cite{Sarica2020}. 

The phase detection computation resulted in a vocabulary of 15,435,308 terms, including 13,730,320 phrases. Since the adopted phrase detection algorithm is purely based on cooccurrence statistics, the detection of some faulty phrases including stopwords such as “the\_”, “a\_”, “and\_”, and “to\_” is inevitable. Therefore, the detected phrases are processed one more time to split the known stopwords from the NLTK \cite{bird2009natural} and USPTO \cite{USPTO} stopwords lists. For example, “an\_internal\_combustion\_engine” is replaced with “an internal\_combustion\_engine”. Then the vocabulary is reduced to 8,641,337 terms, including 6,900,263 phrases.

Next, all the words are represented with their regularized forms to avoid having multiple terms representing the same word or phrase and thus decrease the vocabulary size. This step is achieved by first using a part-of-speech (POS) tagger \cite{Toutanova2007} to detect the type of words in the sentences and lemmatize those words accordingly. For example, if the word “learning” is tagged as a VERB, it would be regularized as “learn” while it would be regularized as “learning” if it is tagged as a NOUN. The lemmatization procedure further decreased the vocabulary to 8,144,852 terms including 6,418,992 phrases. 

As a last step, we removed the words contained in famous NLTK \cite{bird2009natural} and USPTO \cite{USPTO} stopwords lists. The NLTK stopwords list focuses more on general stopwords that can be encountered in daily English language such as “a, an, the, …, he, she, his, her, …, what, which, who, …”, in total 179 words. On the other hand, USPTO stopwords list include words that occur very frequently in patent documents and do not contain critical meaning within patent texts, such as “claim, comprise, … embodiment, … provide, respectively, therefore, thereby, thereof, thereto, …”, in total 99 words. The union of these two lists contains 220 stopwords.

Additionally, we also discarded the words appearing only 1 time in the whole patent database, which leads to a final set of 6,645,391 terms including 5,834,072 phrases.

\subsection{Term Statistics}
\label{sec:termstats}
To identify the frequently occurring words or phrases that carry little information content about engineering and technology, we use four metrics together: 1) direct term frequency (TF), 2) inverse-document frequency (IDF), 3) term-frequency-inverse-document-frequency (TFIDF) and 4) Shannon’s information entropy \cite{Shannon1948}.

We use $f(t)$ to denote direct frequency of term $t$. Consider a corpus $C$ of $P$ patents.

\begin{equation}
TF(t) = \frac{n(t)}{n(p)}
\end{equation}

where $n(p) = \sum\limits_{t}n(t,p)$ is the number of terms in the patent $p$, $n(t) = \sum\limits_{p \in P}n(t,p)$ is total count of term $t$ in all patents. The term frequency is an important indicator of commonality of a term within a collection of documents. Stopwords are expected to have high term frequency.

Inverse-document-frequency (IDF) is calculated as follows

\begin{equation}
IDF(t) = \log\frac{|C|}{DF(t)}
\end{equation}

where $DF(t) = |\{ p \in C : t \in p\}|$  is the number of patents containing term $t$ and $|C|$ represents the number of patents in the database. This metric penalizes the frequently occurring terms and favours the ones occurring in a few documents only. The metric’s lower bound is 0 which refers to the terms that appear in every single document in the database. The upper bound is defined by the terms appearing only in one document, which is $\log |C|$.

Term frequency-inverse-document-frequency (TFIDF) is calculated as follows
\begin{equation}
TFIDF(t) = \frac{1}{DF(t)}\sum\limits_{p}\frac{n(t,p)}{n(p)}\frac{|C|}{DF(t)}
\end{equation}

This metric favours the terms that appear in a few documents, with a considerably high term frequency within the document. If a term appears in many documents, its TFIDF score will be penalized by IDF score due to its commonality. Here, we did not use the traditional IDF metric but removed the log normalizing function to penalize the terms commonly occurring in the entire patent database harder regardless of their in-document (patent) term frequencies. We eventually used the mean of the single document TFIDF scores for each term.

The entropy of term $t$ is calculated as follows. The metric indicates how uneven the distribution of term $t$ is in the corpus $C$.

\begin{equation}
H(t|C) = -\sum\limits_{p} P(p|t)\log P(p|t)
\end{equation}

where $P(p|t)= \frac{n(t,p)}{n(t)}$ is the distribution of term t over patent documents. This indicates how evenly distributed a term is in the patent database. Maximum attainable entropy value for a given collection of documents is basically an even distribution to all patents which leads to $\log|C|$. Therefore, the terms having higher entropy values will contain less information about the patents where they appear, compared to other terms with lower entropy.

We reported the distributions of terms in our corpus according to these four metrics in the Appendix (see Figure \ref{fig:figA1}). The term-frequency distribution has a very long right tail, indicating most of the terms appear a few times in the patent database while some words appear so frequently. Our further tests found that the distribution follows the a power law \cite{Zipf1936,Zipf1949}. By contrast, the distribution by IDF has a long left tail, indicating the existence of a few terms that appears commonly in all patents. The TFIDF distribution also has a long right tail that indicates the existence of highly common terms in each patent and highly strong domain-specific terms dominating a set of patents. Moreover, the long right tail of entropy distribution indicates comparingly few high valued terms that are appearing commonly in the entire database. Therefore, assessing the four metrics together will allow us to detect the stopwords with varied occurrence patterns.

\subsection{Human Evaluation}
\label{sec:humaneval}
We formed 4 different lists of terms sorted by their decreasing TF, increasing IDF, increasing TFIDF, and decreasing entropy. Table \ref{table:A1} in the appendix presents the top ranked 30 terms in respective lists. Then the top 2,000 terms in each of the four lists are used to form a union set of terms. The union only includes 2,305 terms, which indicates that the lists based on four alternative statistic metrics overlap significantly. Then the terms in the union set are evaluated by two researchers with more than 20 years of engineering experience each, in terms of whether a term carries information about engineering and technology, to identify stopwords. The researchers initially achieved an inter-rater reliability of 0.83 \cite{Cronbach1951} and then discussed the discrepancy to reach the consensus on a final list of 62 insignificant terms.

\subsection{Final List}
\label{sec:finallist}
This list, compared to our previous study which identified a list of stopwords \cite{Sarica2020} (see Table \ref{table:A2} in the Appendices) by manually reading 1,000 randomly selected sentences from the same patent text corpus, includes 26 new uninformative stopwords that the previous list did not cover. In the meantime, we also found the previous list contains other 25 stopwords, which are still deemed qualified stopwords in this study. Therefore, we integrate these 25 stopwords from the previous study with the 62 stopwords identified here to derive a final list of 87 stopwords for technical language analysis. The final list is presented in Table \ref{table:table1} together with the NLTK stopwords list and the USPTO stopwords list\footnote{This list can be downloaded from our GitHub repository \url{https://github.com/SerhadS/TechNet}}. It is suggested to apply the three stopwords lists together in technical language processing applications across technical fields.

\begin{table}[h]
\caption{Stopwords lists for technical language processing applications}
\small
\begin{tabular}{cccc|cc|cc}
\hline
\multicolumn{4}{c}{\begin{tabular}[c]{@{}c@{}}NLTK Stopword List {[}10{]}\\ (179 words)\end{tabular}} & \multicolumn{2}{c}{\begin{tabular}[c]{@{}c@{}}USPTO Stopword List {[}18{]}\\ (99 words)\end{tabular}} & \multicolumn{2}{c}{\begin{tabular}[c]{@{}c@{}}This Study\\ (87 words)\end{tabular}} \\
\hline
a        & hadn't   & on         & wasn't     & a             & onto         & able            & others           \\
about    & has      & once       & we         & accordance    & or           & above-mentioned & otherwise        \\
above    & hasn     & only       & were       & according     & other        & accordingly     & overall          \\
after    & hasn't   & or         & weren      & all           & particularly & across          & rather           \\
again    & have     & other      & weren't    & also          & preferably   & along           & remarkably       \\
against  & haven    & our        & what       & an            & preferred    & already         & significantly    \\
ain      & haven't  & ours       & when       & and           & present      & alternatively   & simply           \\
all      & having   & ourselves  & where      & another       & provide      & always          & sometimes        \\
am       & he       & out        & which      & are           & provided     & among           & specifically     \\
an       & her      & over       & while      & as            & provides     & and/or          & straight forward \\
and      & here     & own        & who        & at            & relatively   & anything        & substantially    \\
any      & hers     & re         & whom       & be            & respectively & anywhere        & thereafter       \\
are      & herself  & s          & why        & because       & said         & better          & therebetween     \\
aren     & him      & same       & will       & been          & should       & disclosure      & therefor         \\
aren't   & himself  & shan       & with       & being         & since        & due             & therefrom        \\
as       & his      & shan't     & won        & by            & some         & easily          & therein          \\
at       & how      & she        & won't      & claim         & such         & easy            & thereinto+       \\
be       & i        & she's      & wouldn     & comprises     & suitable     & eg              & thereon          \\
because  & if       & should     & wouldn't   & corresponding & than         & either          & therethrough     \\
been     & in       & should've  & y          & could         & that         & elsewhere       & therewith        \\
before   & into     & shouldn    & you        & described     & the          & enough          & together         \\
being    & is       & shouldn't  & you'd      & desired       & their        & especially      & toward           \\
below    & isn      & so         & you'll     & do            & then         & essentially     & towards          \\
between  & isn't    & some       & you're     & does          & there        & et al           & typical          \\
both     & it       & such       & you've     & each          & thereby      & etc             & typically        \\
but      & it's     & t          & your       & embodiment    & therefore    & eventually      & upon             \\
by       & its      & than       & yours      & fig           & thereof      & excellent       & via              \\
can      & itself   & that       & yourself   & figs          & thereto      & finally         & vice versa       \\
couldn   & just     & that'll    & yourselves & for           & these        & furthermore     & whatever         \\
couldn't & ll       & the        &            & from          & they         & good            & whereas          \\
d        & m        & their      &            & further       & this         & hence           & whereat          \\
did      & ma       & theirs     &            & generally     & those        & he/she          & wherever         \\
didn     & me       & them       &            & had           & thus         & him/her         & whether          \\
didn't   & mightn   & themselves &            & has           & to           & his/her         & whose            \\
do       & mightn't & then       &            & have          & use          & ie              & within           \\
does     & more     & there      &            & having        & various      & ii              & without          \\
doesn    & most     & these      &            & herein        & was          & iii             & yet              \\
doesn't  & mustn    & they       &            & however       & were         & instead         &                  \\
doing    & mustn't  & this       &            & if            & what         & later           &                  \\
don      & my       & those      &            & in            & when         & like            &                  \\
don't    & myself   & through    &            & into          & where        & little          &                  \\
down     & needn    & to         &            & invention     & whereby      & many            &                  \\
during   & needn't  & too        &            & is            & wherein      & may             &                  \\
each     & no       & under      &            & it            & which        & meanwhile       &                  \\
few      & nor      & until      &            & its           & while        & might           &                  \\
for      & not      & up         &            & means         & who          & moreover        &                  \\
from     & now      & ve         &            & not           & will         & much            &                  \\
further  & o        & very       &            & now           & with         & must            &                  \\
had      & of       & was        &            & of            & would        & never           &                  \\
hadn     & off      & wasn       &            & on            &              & often           &  \\
\hline
\end{tabular}
\label{table:table1}
\end{table}
\section{Concluding Remarks}
\label{sec:conclusion}

To develop a comprehensive list of stopwords in engineering and technology-related texts, we mined the patent text database with several statistical metrics from term frequency to entropy together to automatically identify candidate stopwords and use human evaluation to validate, screen and finalize stopwords from the candidates. In this procedure, the automatic data-driven detection of four statistic metrics yield highly overlapping results, and the human evaluations also came with high inter-rater reliability, suggesting evaluator independence. Our final stopwords list can be used as a complementary list to NLTK and USPTO stopwords lists in NLP and text analysis tasks related to technology, engineering design, and innovation.

\bibliographystyle{unsrt}  
\bibliography{main}  

\begin{thebibliography}{10}

\bibitem{Chang2015}
Danni Chang and Chun-hsien Chen.
\newblock {Product concept evaluation and selection using data mining and
  domain ontology in a crowdsourcing environment}.
\newblock {\em Advanced Engineering Informatics}, 29(4):759--774, oct 2015.

\bibitem{Zhang2014}
Yi~Zhang, Alan~L. Porter, Zhengyin Hu, Ying Guo, and Nils~C. Newman.
\newblock {"Term clumping" for technical intelligence: A case study on
  dye-sensitized solar cells}.
\newblock {\em Technological Forecasting and Social Change}, 85:26--39, 2014.

\bibitem{Grawe2017}
Mattyws~F Grawe, Claudia~A Martins, and Andreia~G Bonfante.
\newblock {Automated Patent Classification Using Word Embedding}.
\newblock In {\em 2017 16th IEEE International Conference on Machine Learning
  and Applications (ICMLA)}, pages 408--411. IEEE, dec 2017.

\bibitem{Liu2020}
Qiyu Liu, Kai Wang, Yan Li, and Ying Liu.
\newblock {Data-driven Concept Network for Inspiring Designers' Idea
  Generation}.
\newblock {\em Journal of Computing and Information Science in Engineering},
  pages 1--39, 2020.

\bibitem{Blanchard2007}
Antoine Blanchard.
\newblock {Understanding and customizing stopword lists for enhanced patent
  mapping}.
\newblock {\em World Patent Information}, 29(4):308--316, 2007.

\bibitem{Gerlach2019}
Martin Gerlach, Hanyu Shi, and Luis~A.Nunes Amaral.
\newblock {A universal information theoretic approach to the identification of
  stopwords}.
\newblock {\em Nature Machine Intelligence}, 2019.

\bibitem{Tsz-WaiLo2005}
Rachel {Tsz-Wai Lo}, Ben He, and Iadh Ounis.
\newblock {Automatically Building a Stopword List for an Information Retrieval
  System}.
\newblock In {\em 5th Dutch-Belgium Information Retrieval Workshop}, Utrecht,
  2005.

\bibitem{Fox1989}
Christopher Fox.
\newblock {A stop list for general text}.
\newblock {\em ACM SIGIR Forum}, 24(1-2):19--21, sep 1989.

\bibitem{Wilbur1992}
W.~John Wilbur and Karl Sirotkin.
\newblock {The automatic identification of stop words}.
\newblock {\em Journal of Information Science}, 18(1):45--55, 1992.

\bibitem{bird2009natural}
Steven Bird, Ewan Klein, and Edward Loper.
\newblock {\em {Natural language processing with Python: analyzing text with
  the natural language toolkit}}.
\newblock O'Reilly Media, Inc., 2009.

\bibitem{Kucera1969}
Henry Ku{\v{c}}era and Winthrop~Nelson Francis.
\newblock {Computational analysis of present-day American English}.
\newblock {\em International Journal of American Linguistics}, 35(1):71--75,
  1969.

\bibitem{Montemurro2010}
Marcelo~A. Montemurro and Dami{\'{a}}n~H. Zanette.
\newblock {Towards the quantification of the semantic information encoded in
  written language}.
\newblock {\em Advances in Complex Systems}, 13(2):135--153, 2010.

\bibitem{Sarica2020}
Serhad Sarica, Jianxi Luo, and Kristin~L. Wood.
\newblock {TechNet: Technology semantic network based on patent data}.
\newblock {\em Expert Systems with Applications}, 142, 2020.

\bibitem{Seki2005}
Kazuhiro Seki and Javed Mostafa.
\newblock {An application of text categorization methods to gene ontology
  annotation}.
\newblock {\em SIGIR 2005 - Proceedings of the 28th Annual International ACM
  SIGIR Conference on Research and Development in Information Retrieval}, pages
  138--145, 2005.

\bibitem{Crow2004}
Dan Crow and John Desanto.
\newblock {A hybrid approach to concept extraction and recognition-based
  matching in the domain of human resources}.
\newblock {\em Proceedings - International Conference on Tools with Artificial
  Intelligence, ICTAI}, (Ictai):535--539, 2004.

\bibitem{Makrehchi2008}
Masoud Makrehchi and Mohamed~S. Kamel.
\newblock {Automatic extraction of domain-specific stopwords from labeled
  documents}.
\newblock {\em Lecture Notes in Computer Science (including subseries Lecture
  Notes in Artificial Intelligence and Lecture Notes in Bioinformatics)}, 4956
  LNCS:222--233, 2008.

\bibitem{Mikolov2013}
Tomas Mikolov, Kai Chen, Greg Corrado, and Jeffrey Dean.
\newblock {Distributed Representations of Words and Phrases and their
  Compositionality}.
\newblock In {\em Advances in Neural Information Processing Systems (NIPS) 26},
  pages 1--9, 2013.

\bibitem{USPTO}
USPTO.
\newblock {Stopwords, USPTO Full-Text Database}.

\bibitem{Toutanova2007}
Kristina Toutanova and Christopher~D. Manning.
\newblock {Enriching the knowledge sources used in a maximum entropy
  part-of-speech tagger}.
\newblock In {\em EMNLP '00 Proceedings of the 2000 Joint SIGDAT conference on
  Empirical methods in natural language processing and very large corpora},
  pages 63--70, 2007.

\bibitem{Shannon1948}
C.~E. Shannon.
\newblock {A Mathematical Theory of Communication}.
\newblock {\em Bell System Technical Journal}, 27(3):379--423, jul 1948.

\bibitem{Zipf1936}
George~Kingsley Zipf.
\newblock {\em {The Psychobiology of Language}}.
\newblock Routledge, London, 1936.

\bibitem{Zipf1949}
George~Kingsley Zipf.
\newblock {\em {Human Behavior and the Principle of Least Effort.}}
\newblock Addison-Wesley, New York, 1949.

\bibitem{Cronbach1951}
Lee~J. Cronbach.
\newblock {Coefficient alpha and the internal structure of tests}.
\newblock {\em Psychometrika}, 16(3):297--334, sep 1951.

\end{thebibliography}

\appendix
\appendixpage
\setcounter{figure}{0}
\renewcommand\thefigure{A\arabic{figure}}
\setcounter{table}{0}
\renewcommand{\thetable}{A\arabic{table}}

\begin{table}[h]
\centering
\caption{Top 30 terms for term-frequency, IDF, TFIDF and entropy}
\begin{tabular}{ccccc}
\hline
   & Term-Frequency & IDF        & TFIDF      & Entropy    \\
\hline
1  & method         & method     & include    & method     \\
2  & first          & include    & method     & include    \\
3  & include        & one        & one        & one        \\
4  & second         & first      & comprise   & form       \\
5  & form           & form       & form       & first      \\
6  & one            & comprise   & system     & comprise   \\
7  & system         & system     & first      & system     \\
8  & plurality      & second     & least      & second     \\
9  & device         & plurality  & second     & apparatus  \\
10 & comprise       & apparatus  & apparatus  & plurality  \\
11 & apparatus      & device     & plurality  & least      \\
12 & least          & least      & receive    & disclose   \\
13 & least\_one     & disclose   & disclose   & device     \\
14 & may            & receive    & device     & receive    \\
15 & connect        & may        & connect    & may        \\
16 & process        & least\_one & may        & connect    \\
17 & control        & connect    & position   & least\_one \\
18 & portion        & control    & control    & control    \\
19 & receive        & process    & least\_one & process    \\
20 & position       & position   & portion    & position   \\
21 & mean           & portion    & base       & base       \\
22 & surface        & base       & determine  & portion    \\
23 & say            & surface    & generate   & surface    \\
24 & base           & determine  & make       & determine  \\
25 & disclose       & generate   & surface    & make       \\
26 & configure      & make       & within     & generate   \\
27 & determine      & mean       & process    & relate     \\
28 & generate       & produce    & accord     & produce    \\
29 & substrate      & configure  & end        & configure  \\
30 & signal         & relate     & allow      & within    \\
\hline
\end{tabular}
\label{table:A1}
\end{table}

\begin{table}
\centering
\caption{The stopwords identified in the previous study. * indicates that the term is also identified in the current study. + indicates that the term is a stopword as defined in the current study. Rest of the terms are no longer considered as stopwords as defined in the current study.}
\begin{tabular}{cccc}
\hline
able*            & etc*         & one another       & therethrough* \\
above-mentioned+ & eventually+  & otherwise*        & therewith*    \\
already*         & finally*     & possibly          & towards*      \\
always*          & furthermore* & rather*           & typical+      \\
and/or*          & he/she+      & remarkably+       & via*          \\
anything+        & hence*       & significantly+    & vice versa+   \\
anywhere+        & him/her+     & simply*           & whatever+     \\
better*          & his/her+     & sometimes+        & whereat+      \\
disclosure+      & instead*     & straight forward+ & wherever+     \\
easily*          & may*         & substantially     & whether*      \\
eg*              & meanwhile+   & therebetween*     & whose*        \\
either*          & might+       & therefor*         & within*       \\
elsewhere+       & moreover+    & therefrom*        & without*      \\
enough+          & must*        & therein*          & wrt           \\
especially*      & often+       & thereinto+        & yet*          \\
et al+           & one          & thereon*          &        \\
\hline
\end{tabular}
\label{table:A2}
\end{table}

\begin{figure}[]
  \centering
  \includegraphics[scale=0.70]{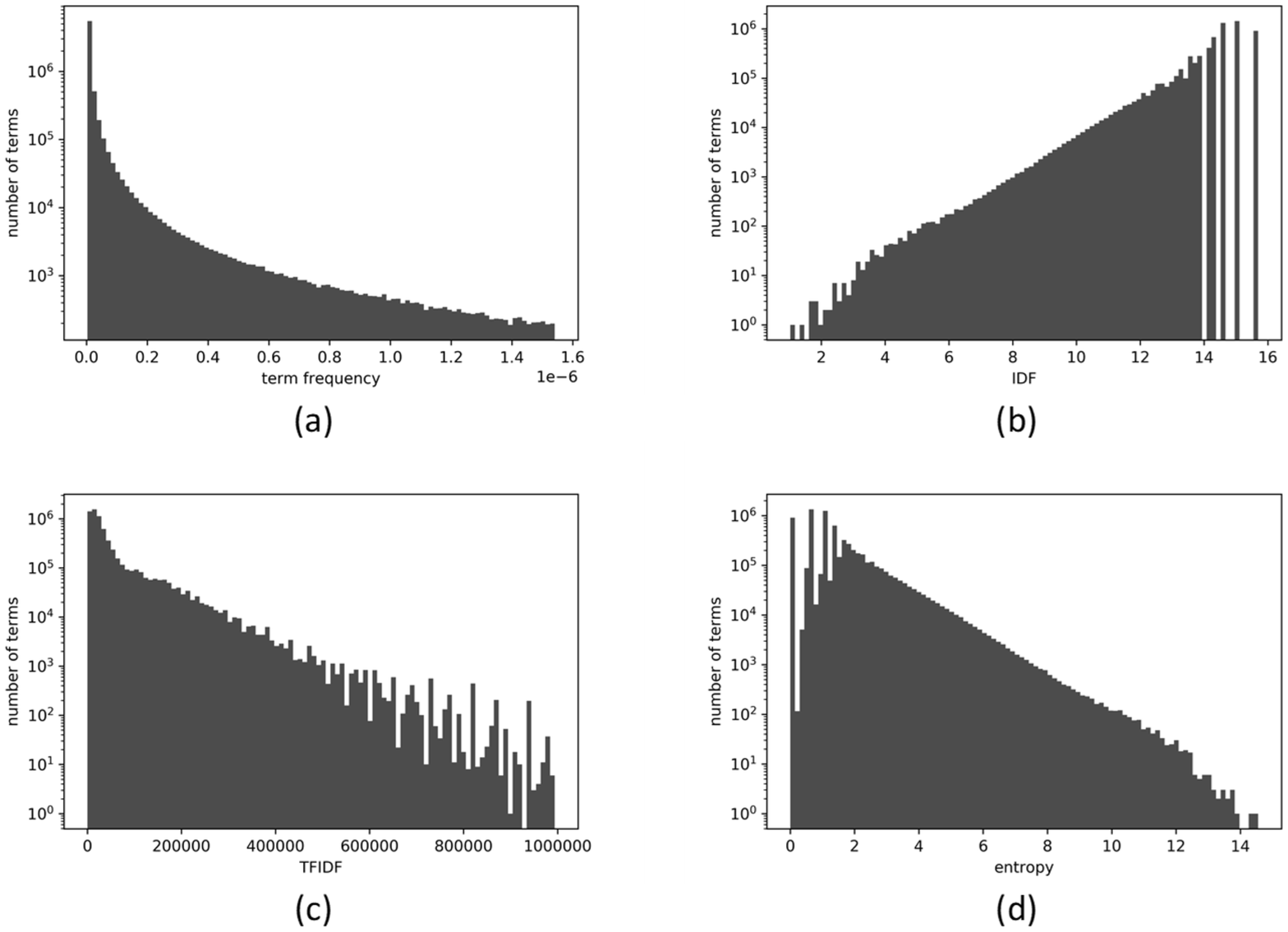}
  \caption{Distribution of terms by (a) term-frequency, (b) IDF, (c) TFIDF and (d) entropy. Term-frequency and TFIDF histograms arbitrarily filtered (term-count$<=$1000, TFIDF score$<=10^6$) for visualization purposes. In fact, they have longer right tails.}
  \label{fig:figA1}
\end{figure}
\end{document}